 \documentclass[smallabstract,smallcaptions]{dccpaper}

\usepackage{epsfig}
\usepackage{amsmath}
\usepackage{amssymb}
\usepackage{color}
\usepackage{url}
\usepackage[hyperfootnotes=false]{hyperref}
\newlength{\figurewidth}
\newlength{\smallfigurewidth}

\begin{document}

\title
{\large
\textbf{Reducing latency and bandwidth for video streaming using keypoint extraction and digital puppetry}
}

\author{%
Roshan Prabhakar$^{1}$, Shubham Chandak$^{2}$, Carina Chiu$^{3,*}$, Renee Liang$^{4,*}$,\\ Huong Nguyen$^{5,*}$, Kedar Tatwawadi$^{2}$, 
and Tsachy Weissman$^{2}$\\[0.5em]
$^{1}$Fremont High School, Sunnyvale, CA, USA\\
$^{2}$Department of Electrical Engineering, Stanford University, Stanford, CA, USA\\
$^{3}$Prospect High School, Saratoga, CA, USA\\
$^{4}$Northwood High School, Irvine, CA, USA\\
$^{5}$Leesville Road High School, Raleigh, NC, USA\\
Contact: \url{roshan@aphrabhakar.com}, \url{schandak@stanford.edu}\\
}

\maketitle
\thispagestyle{empty}
\let\thefootnote\relax\footnote{* These authors contributed to the initial prototype of this work while participating in the STEM to SHTEM high school summer internship program at Stanford University.}

\begin{abstract}
COVID-19 has made video communication one of the most important modes of information exchange. While extensive research has been conducted on the optimization of the video streaming pipeline, in particular the development of novel video codecs, further improvement in the video quality and latency is required, especially under poor network conditions. This paper proposes an alternative to the conventional codec through the implementation of a keypoint-centric encoder relying on the transmission of keypoint information from within a video feed. The decoder uses the streamed keypoints to generate a reconstruction preserving the semantic features in the input feed. Focusing on video calling applications, we detect and transmit the body pose and face mesh information through the network, which are displayed at the receiver in the form of animated puppets. Using efficient pose and face mesh detection in conjunction with skeleton-based animation, we demonstrate a prototype requiring lower than 35 kbps bandwidth, an order of magnitude reduction over typical video calling systems. The added computational latency due to the mesh extraction and animation is below 120ms on a standard laptop, showcasing the potential of this framework for real-time applications. The code for this work is available at \url{https://github.com/shubhamchandak94/digital-puppetry/}.

\end{abstract}

\Section{Introduction}

Over the years, there has been a significant amount of work on the development and optimization of novel video codecs which are a key element in the streaming pipeline. Numerous codecs have been developed in order to reduce the amount of streamed data while maintaining as much information as possible in the data stream. In a typical system, first the data is read from a video feed and compressed. The compressed data is sent over a network to the receiving end, where a decoding algorithm reconstructs a representation of the original feed from the streamed data. Most of the codecs are lossy, i.e., the reconstruction process does not recreate the original feed; instead the reconstruction is sufficiently close to the original video under some measure of distortion. The lossy compression algorithms attempt to utilize the fact that not all the information contained within a video frame is equally important and prioritize the preservation of more important aspects of a feed over others in the compression/decompression process. As an example, the human eye is more sensitive to brightness than to color \cite{winkler2001visionchroma}, and hence the compressor can use a coarser representation for the color without a noticeable impact for the human viewer. 

Despite these advances, the current systems can suffer from glitches, lags and poor video quality, especially under unreliable network conditions \cite{fouladi2018salsify}. These issues are particularly problematic in certain delay-sensitive applications such as the performing arts \cite{ubik2020ultra}. Note that in many applications, it is not necessary to reconstruct an approximation of the original video as long as the semantically relevant features are preserved. This observation forms the basis of our approach as described below.

\SubSection{Our contributions}
In this work, we propose an alternative approach that attempts to drastically reduce the bandwidth consumption by selectively transmitting certain \emph{keypoints}. The idea is that these keypoints carry the most important information in the input feed, and can be used at the receiver to generate a reconstruction that captures the essential features of the input rather than attempting to approximate the original video at the pixel level. We focus here on the video calling application, where the facial expression and the general body pose can be considered more important than the background information. Keypoints related to the general facial structure and the body of the subject are extracted and transmitted as an encoded representation of the originating spatial feed. Thus, the encoding process becomes the scraping of keypoint related data, and the decoding process becomes a reconstruction using the scraped data, in the form of some animation. 

One possible concern with such an approach is that it shifts the burden from the communication bandwidth to increased computation at the sender and receiver. The keypoint extraction and animation process can lead to higher latency and make the approach unsuitable for real-time applications. However, as demonstrated in this work, it is possible to achieve an order of magnitude reduction in bandwidth consumption over standard video streaming systems while maintaining sufficiently low latency to enable video calling. This is achieved using state-of-the-art pose and face mesh extraction methods relying on optimized neural network models. We use a skeleton-based animation system for generating the reconstruction, although the framework is extensible to more sophisticated and realistic reconstructions. The  code for the proposed framework is available on GitHub and can be useful as a starting point for further investigation into this novel streaming framework. 

\SubSection{Related work}

There has been multiple decades of research on video codecs for various applications \cite{richardson2002video}, including the most recent AV1 \cite{de2018av1} and VVC \cite{vvc} codecs. In addition, there has been significant recent interest in improving these codecs using machine learning approaches, either using end-to-end approaches or focusing on specific parts of the coding pipeline \cite{liu2020deep}. This has included work on directly optimizing the schemes for human perception and designing new distortion metrics such as VMAF \cite{li2018vmaf} to improve upon traditional metrics such as RMSE and MS-SSIM \cite{wang2003multiscale}.

More closely related to the current work, there have been numerous papers on face detection/mesh extraction \cite{bazarevsky2019blazeface,kartynnik2019realfacemesh,bulat2017far,grishchenko2020attention} and on body pose tracking \cite{bazarevsky2020blazepose,sun2019deep,papandreou2018personlab}, focusing on both 3D and 2D meshes and typically based on neural network models. On the animation side, several current smartphone platforms allow for generating animation based on face detection, for example, Apple Memoji and Snapchat filters. In addition, platforms such as Adobe character animator \cite{adobe} allow a performer to control an animated puppet based on the facial expression and pose. Finally, there has been recent interest in the generation of realistic looking videos and images \cite{liu2020generative}, in particular Deepfake models that use autoencoder or GANs (generative adversarial network) to generate fake realistic videos with human subjects \cite{dolhansky2020deepfake}. Note that these papers were focused on detection or reconstruction tasks and did not involve the streaming aspect, which is the key contribution of this work. There was some work on video compression and reconstruction based on facial landmarks in \cite{eisert1998analyzing,eisert2003mpeg}, which showed the promise of extremely low bitrates, but did not demostrate real-time conferencing capabilities.

While this manuscript was in preparation, NVIDIA independently announced the NVIDIA Maxine project which uses a very similar approach as the current work and aims to reduce the bandwidth consumption in video conferencing \cite{nvidiablog}. The main difference is that they use more powerful GAN-based reconstruction to produce realistic looking videos at the receiver, and thus require GPUs to enable real-time performance. The reported bandwidth consumption is similar to this work, although many details are still unclear. We note that a preliminary version of our work was published online as a blog post \cite{informaticistsblog} more than a month before the NVIDIA announcement.

\begin{figure}[htbp]
    \centering
    \includegraphics[width=\textwidth]{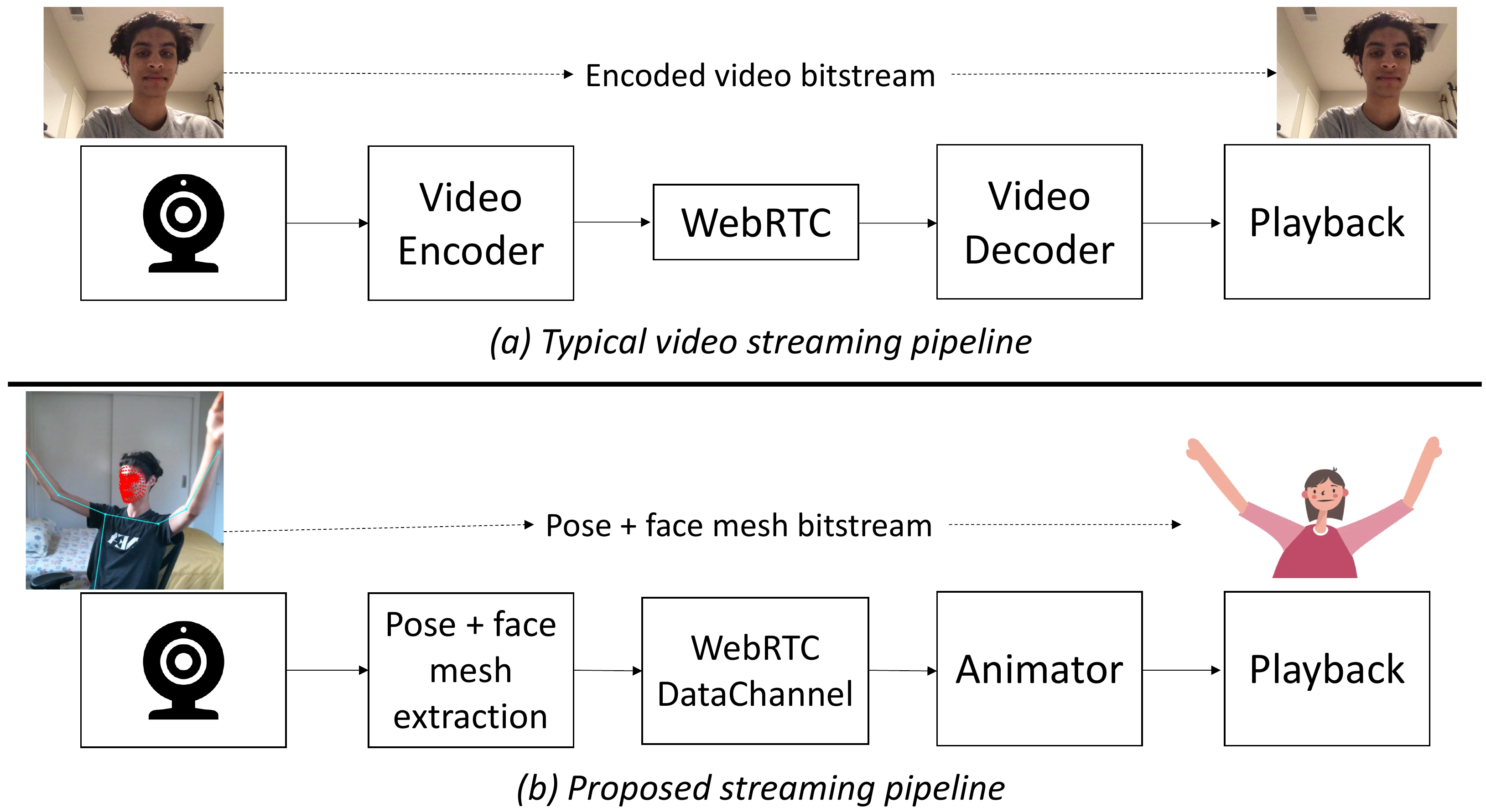}
    \caption{Comparison of (a) typical video streaming pipeline with (b) proposed streaming pipeline. In the typical system, the input video is encoded using video codecs and sent to the receiver which decodes it in the form of a lossy reconstruction that preserves most of the video features at a pixel level. In the proposed system, the body pose and face mesh are extracted at the encoder and transmitted through a general-purpose WebRTC DataChannel. This mesh information is used at the decoder to display animated puppets during the playback.}
    \label{fig:pipeline}
\end{figure}

\Section{Methods}
Figure \ref{fig:pipeline} shows the proposed framework and compares it to the traditional video streaming pipeline. In the proposed framework, the video encoder is replaced by a pose and face mesh extractor. Rather than transmitting the encoded video through a video-specific channel, we transmit the mesh points and confidence scores through a general data channel. The decoder uses the received mesh points to generate an animated video with a customizable character/puppet. We next discuss the two major components we used to build the prototype system: (i) WebRTC \cite{loreto2014realwebrtc} for the communication, and (ii) Pose Animator \cite{poseanimator} for the mesh extraction and animation.

\SubSection{WebRTC}
Web(R)eal(T)ime(C)ommunication \cite{loreto2014realwebrtc} is an open web standard for real-time peer-to-peer communication on the web. WebRTC provides an easy-to-use JavaScript API for the various steps involved in the streaming process. The streaming involves two key steps, the first being the exchange of SDP addresses between peers which can be done over any mechanism. In our prototype, we use a WebSocket \cite{fette2011websocket} based signaling server that allows this exchange. Once the exchange is done, a direct connection is set up between the two peer nodes. For testing and demonstrating the conventional video streaming pipeline, we use the high-level API for the RTCPeerConnection interface that is optimized for audio/video data. For the proposed framework which transmits the keypoint information through the channel, we use the RTCDataChannel interface that allows transfer of arbitrary data between peers. 

\begin{figure}[htbp]
    \centering
    \includegraphics[width=0.5\textwidth]{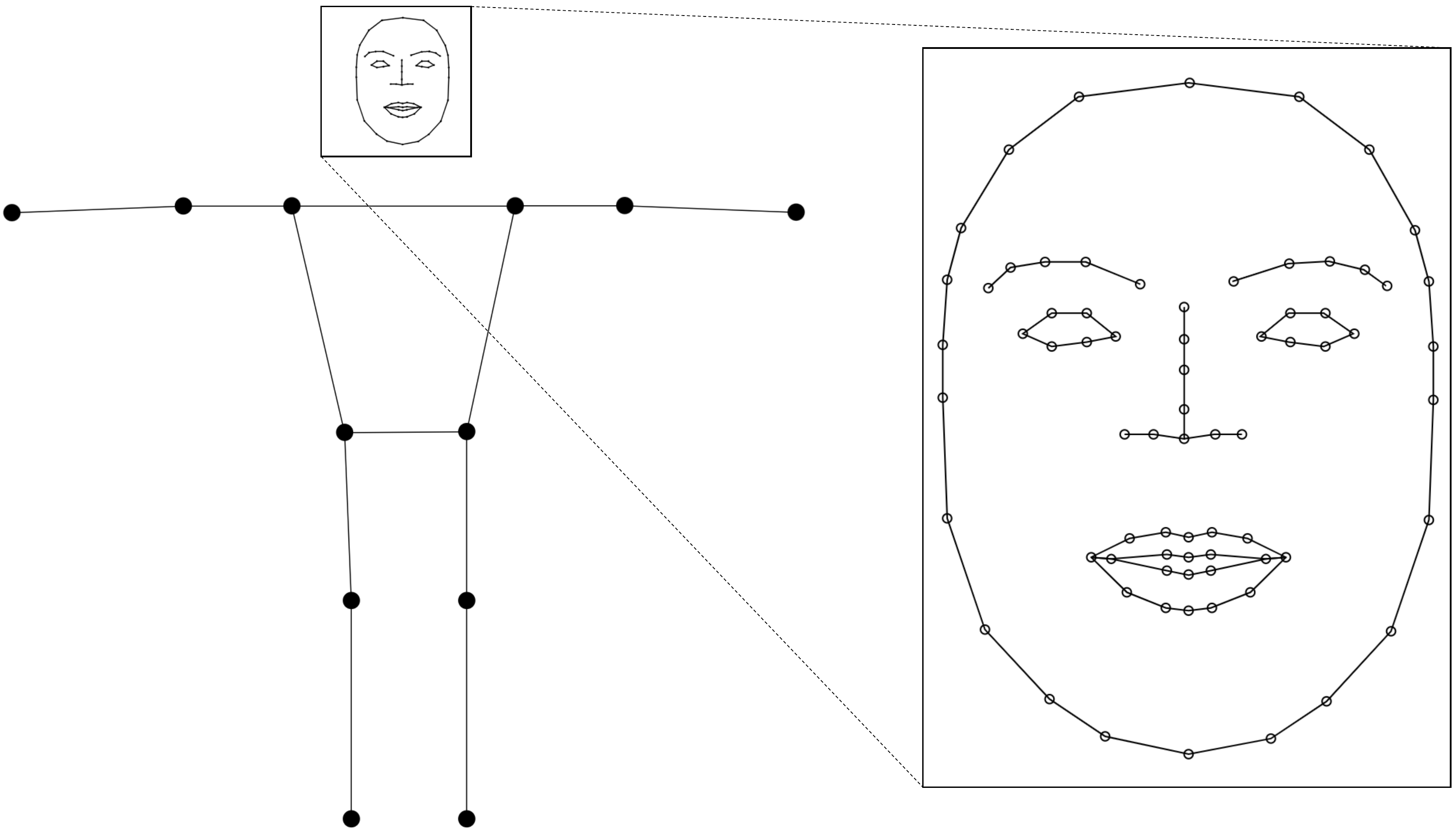}
    \caption{The skeleton points based on Pose Animator \cite{poseanimator} used for the animation in the proposed framework, with the face mesh shown in the inset. The pose structure has 17 keypoints (including 5 on the face that are not shown here), while the facemesh has 73 keypoints.}
    \label{fig:keypoints}
\end{figure}

\SubSection{Pose Animator}
Pose Animator \cite{poseanimator} is an open source project that tracks a person's face and body pose through a webcam and maps the keypoints to generate a skeleton-based animation. Pose Animator combines two main technologies: PoseNet \cite{papandreou2018personlab} (library allowing extraction of pose information pertaining to a human body) and FaceMesh \cite{kartynnik2019realfacemesh} (library for extraction of facemesh from a video feed). Both these libraries focus on efficiency and can run in real-time on a laptop without specialized hardware like GPUs. For the animation, Pose Animator projects the pose and facemesh information onto a provided SVG (Scalable Vector Graphics) skeleton using a rigging algorithm for this task. The animation also uses confidence scores from the mesh detection and motion stabilization to improve the animation quality and stability. We utilize this project for implementing the keypoint extraction and animation components in the proposed pipeline.  

\begin{figure}[htbp]
    \centering
    \includegraphics[width=\textwidth]{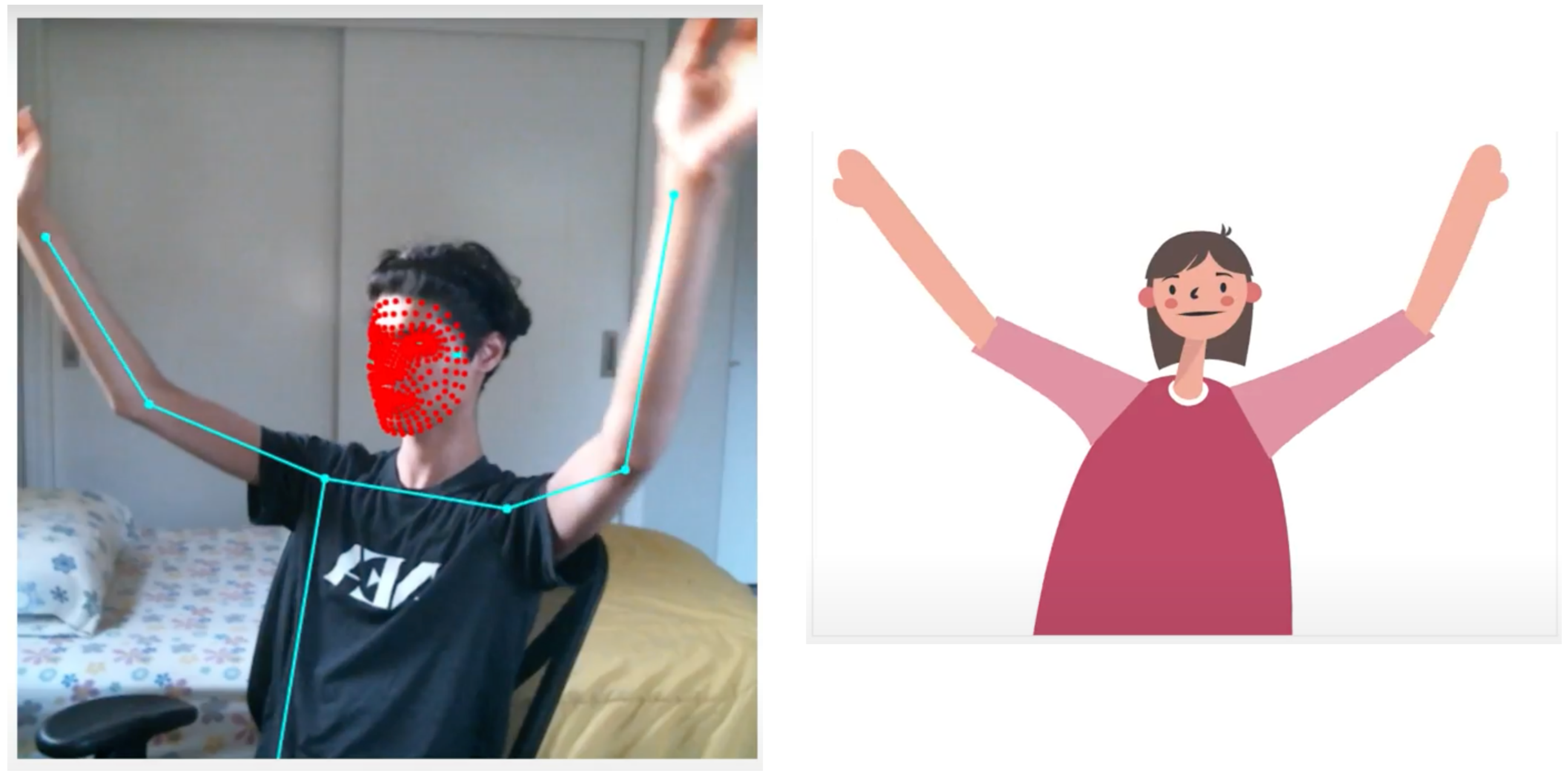}
    \caption{Demonstration of the proposed framework under ideal network conditions. The left panel shows the input video feed with the pose and face mesh superimposed on the video. The right panel shows the generated animation at the receiver. The rotated view is due to the two-camera setup. The full demo video showing the bandwidth and latency statistics is available at \url{https://youtu.be/70rH_jYf_LU}.}
    \label{fig:proposed}
\end{figure}

The animation skeleton is shown in Figure \ref{fig:keypoints}, with the face mesh shown in an inset. The pose skeleton is comprised of the following 17 key points, where each keypoint is represented by a confidence score and two positions (x and y). The entire pose is additionally coupled with a cumulative confidence score. The facemesh is composed of 73 keypoints, with x and y positions for each keypoint and a cumulative confidence score. Note that the detected facemesh has substantially more points than this (as seen in the left panel of Figure \ref{fig:proposed}), but we only transmit the points used in the animation. These positions and confidence scores are transmitted as binary values (16-bit integer or a 32-bit float depending on the required precision). The net data transferred per frame is around 832 bits for the pose and 2368 bits for the face mesh, with a total of 3200 bits/frame. The current prototype does not use any spatial or temporal compression of these keypoints. Despite this, the proposed framework requires around 4x fewer bits per frame compared to a low-quality video stream with a frame rate of 15 fps and bandwidth of 200 kbps \cite{agora}.  

\Section{Experiments and results}
\SubSection{Experimental Setup}

We tested both the conventional and the proposed streaming pipelines with the prototype available at \url{https://github.com/shubhamchandak94/digital-puppetry}. For the proposed keypoint-based framework, the two peers initially connect to a mutual signalling WebSocket interface hosted on a public server to initiate the WebRTC connection. Once the connection has been established, streaming may begin, with the architecture supporting both one-way and two-way streaming. For the sake of logistical simplicity, the client nodes are connected to the same local area network. To test the applicability of this solution in general use cases, both machines are laptops without specialized GPU hardware (MacBooks using Chrome browser). Physically, both laptops are situated next to each other, with roughly 2 feet of distance separating them from each other. Both laptops face the same human subject, thus the corresponding renderings would be of the same subject, just from different angles corresponding to each laptop’s perspective.

The following trial statistics are sampled during the testing process:
\begin{itemize}
    \item One-way bandwidth consumption at any given time (bitrate) 
    \item Net latency (video capture to animation render)
    \item Keypoint extraction latency
    \item Transmission latency
    \item Animation render latency
\end{itemize}

For the conventional streaming model, we performed a similar test. WebRTC does not provide an API to extract latency statistics from a video-specific stream between two RTCPeerConnections, but we are able to measure the stream’s bandwidth consumption at any given time. The implementation of the conventional stream shows the video capture and the video render on the same machine. Data is sent from the originating feed through a virtual channel and is rendered on the same machine. We also implemented the ability to throttle the bandwidth resource available on this virtual channel.

We note that this system is not nearly as robust as one that would be needed in a production environment, it is simply a proof of concept to demonstrate the main ideas. Most of the tests are performed under relatively ideal network conditions (local area network), but this doesn't affect the most important metrics which are the bitrate and the extraction/render latency. In addition, the current system is focused on the video streaming aspect and does not support audio which is not enabled in either streaming method.

\begin{table}[htbp]
\begin{center}
{
\begin{tabular}{cc}
\hline
Measurement type & Typical range \\
\hline
Bitrate &  25-35 kbps              \\
\hline
Net latency & 140-190 ms           \\
Extraction latency &  60-100 ms\\
Transmission latency & 40-60 ms   \\
Render latency &  10-15 ms     \\
\hline
\end{tabular}}
\caption{\label{table:latency_measurements}%
Typical ranges for the bitrate and latency measurements for the proposed framework tested under ideal network conditions.}
\end{center}
\end{table}

\begin{figure}[htbp]
    \centering
    \includegraphics[width=\textwidth]{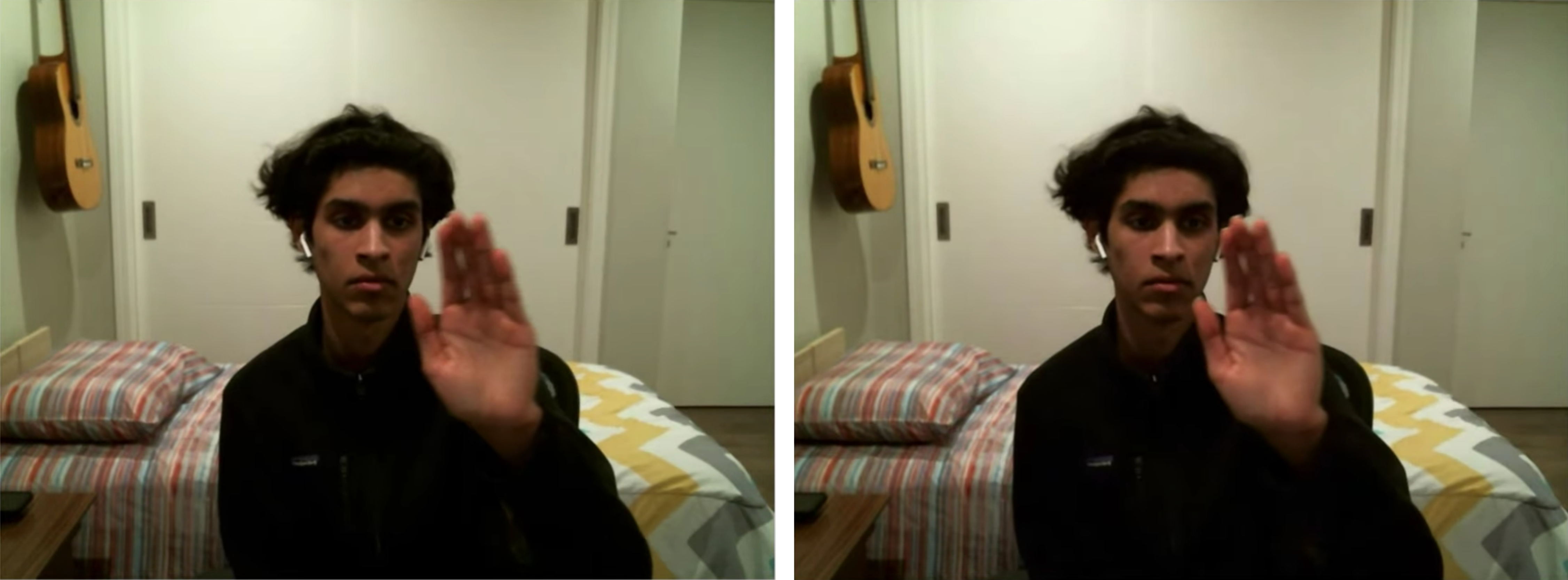}
    \caption{Demonstration of the conventional framework without any bandwidth constraint under ideal network conditions, requiring around 1700 kbps transmission bandwidth. The left and right panels show the input and received video feeds, respectively, with close to zero latency. The full demo video showing the bandwidth statistics is available at \url{https://youtu.be/gmF6oLm7Fkc}.}
    \label{fig:traditional_unlimited}
\end{figure}

\SubSection{Results}

Figure \ref{fig:proposed} shows a demonstration of the proposed framework (with the full video available at \url{https://youtu.be/70rH_jYf_LU}). The corresponding statistics are shown in Table \ref{table:latency_measurements}. We can see the bandwidth consumption is around 25-35 kpbs which is an order of magnitude lower than that for typical video calling systems \cite{agora}. With this relatively small bandwidth requirement the setup demonstrates the potential for real-time connectivity even in the unreliable network conditions. 

The net latency is around 140-190 ms, which is generally considered to be tolerable for video calling applications \cite{zoomlatency}. We believe that further optimization in the extraction and render latency can be achieved with further optimizations and possibly with the use of GPUs or multithreading. A related aspect that needs further optimization is the frame rate which was slightly low (10-11 frames/s), mainly due to the extraction latency. Again, it should be straightforward to improve this using GPUs or multithreading. We found the transmission latency to be highly variable between different trials on the local area network (measurements ranged from $\sim$15ms to $\sim$50ms), and further experimentation under different network conditions can help better understand the benefits of the proposed scheme for this metric.

\begin{figure}[htbp]
    \centering
    \includegraphics[width=\textwidth]{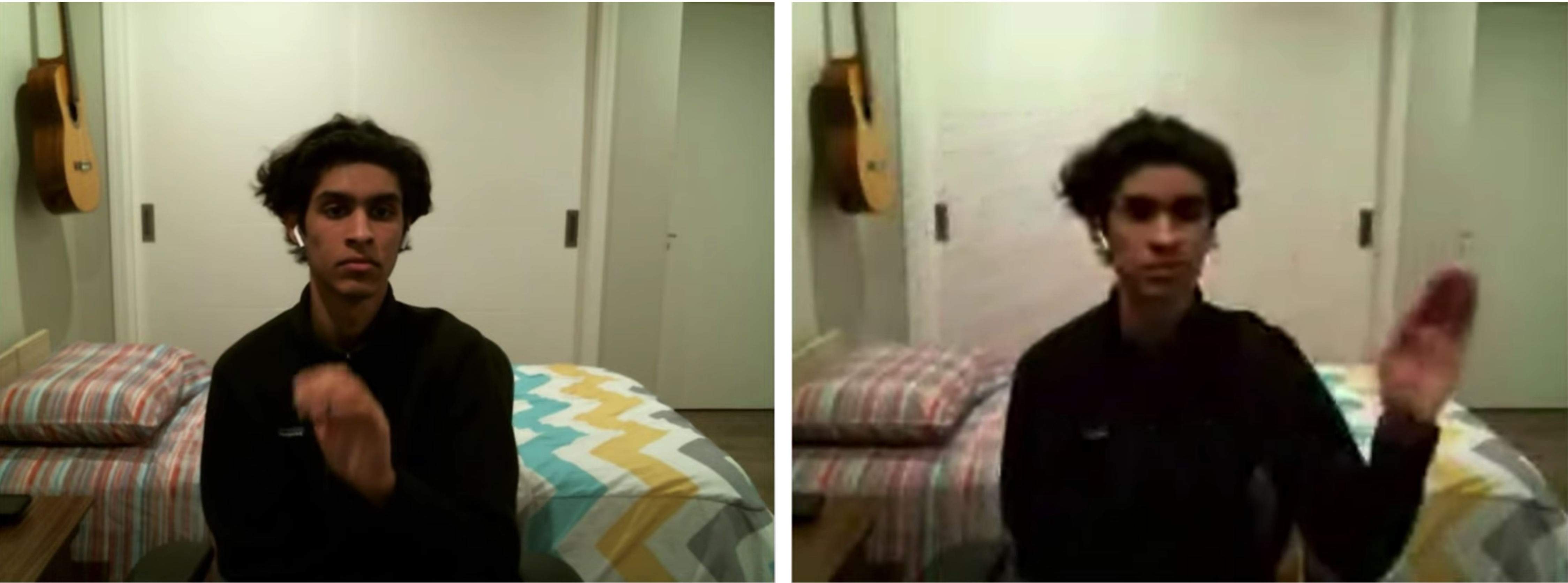}
    \caption{Demonstration of the conventional framework with bandwidth constraint of 35 kbps (to match the bandwidth consumption of the proposed framework). The left and right panels show the input and received video feeds, respectively. We can observe the latency in the hand motion and the blurred video quality at the receiver due to the limited bandwidth. The full demo video showing the bandwidth statistics is available at \url{https://youtu.be/gmF6oLm7Fkc}.}
    \label{fig:traditional_limited}
\end{figure}

Figures \ref{fig:traditional_unlimited} and \ref{fig:traditional_limited}  show the conventional streaming framework with no bandwidth constraint and with a bandwidth constraint of 35 kbps, respectively. The complete demonstration video is available at \url{https://youtu.be/gmF6oLm7Fkc}. For the unconstrained setting, the performance was good, but the bitrate was around 1700 kbps, which is around 50x higher than that required for the proposed system. When the bandwidth is constrained to 35 kbps (which matches the bitrate for the keypoint stream), we observe significant lags and blurring especially when the subject is moving.

\Section{Discussion and conclusion}
We observe that the novel streaming pipeline can reduce the bandwidth consumption to around 35 kbps which is an order of magnitude lower than the conventional streaming systems. In addition, the latency introduced by the keypoint extraction and animation rendering stages is sufficiently low ($\sim$120ms) to enable real-time communication without the need for specialized hardware. We also observe that restricting the conventional model to comparable bandwidth consumption leads to noticeable reduction in video quality as well as increased latency. Clearly, the proposed system is ideal for bandwidth-constrained environments, particularly in areas with unreliable internet access. Another application we are currently exploring is puppet-based live performances in virtual theater to reduce the actor-to-actor latency. This scheme also has interesting implications for privacy, for example, one could envision communicating the audience reactions (in the form of keypoints) to a sports match or a theater performance without needing to actually share the video itself. 

Although this implementation provides a proof of concept for the underlying idea, further work is needed to implement a practical alternative to the current systems. In addition to building a robust production-ready system and more extensive testing, there is a need for further reduction in the keypoint extraction and animation rendering latency, improvement in the animation quality, and lossless or lossy compression of the transmitted keypoints by exploiting the inherent spatiotemporal redundancy. The techniques from \cite{eisert2003mpeg} can be utilized to achieve bitrates as low as 1 kbps and to perform a systematic exploration of the rate-distortion tradeoff. While the current animation pipeline provides an easy-to-use and lightweight mechanism for projecting the keypoints, we believe 3D character animation or deep learning-based realistic human reconstruction (e.g., deepfake) can provide a more satisfying user experience. We note that improvements in the keypoint extraction and animation require optimized implementations to allow real-time communication, possibly with the help of specialized hardware such as a GPU. With the rapid improvement of hardware capabilities in mobiles and personal computers, this is unlikely to be a major obstacle. As evidenced by the recent announcement of the NVIDIA Maxine project \cite{nvidiablog}, we believe that these hurdles are surmountable and these ideas can be translated into a practical system that provides immense gains over the conventional methods. 

Finally, we note that the core idea of extracting and transmitting keypoints is applicable beyond the physical keypoints (pose and face mesh) utilized in this work. In general, keypoints can be more abstract quantities, for example, the emotion of the person reconstructed as an emoji or the intermediate outputs of a neural network (e.g., an autoencoder) used to produce an appropriate reconstruction at the receiver. The type of information communicated can be chosen depending on the requirements of the application, the network constraints and the computational capabilities of the nodes.  

\Section{Acknowledgements}
We thank the Stanford Compression Forum and the STEM to SHTEM high school internship program for providing us the opportunity to work on this project. We also thank Michael Rau and Zachary Hoffman for fruitful discussions.

\Section{References}
\bibliographystyle{IEEEbib}
\bibliography{main}

\end{document}